# How complicated must an optical component be?

**David A. B. Miller**[*]
*Ginzton Laboratory, Stanford University, 348 Via Pueblo Mall,*
[*]*Corresponding author: dabm@ee.stanford.edu*

We analyze how complicated a linear optical component has to be if it is to perform one of a range of functions. Specifically, we devise an approach to evaluating the number of real parameters that must be specified in the device design or fabrication, based on the singular value decomposition of the linear operator that describes the device. This approach can be used for essentially any linear device, including space-, frequency-, or time-dependent systems, in optics or in other linear wave problems. We analyze examples including spatial mode converters and various classes of wavelength demultiplexers. We consider limits on the functions that can be performed by simple optical devices such as thin lenses, mirrors, gratings, modulators, and fixed optical filters, and discuss the potential for greater functionalities using modern nanophotonics.

## 1. Introduction

With the recent rapid growth in our ability to fabricate novel and complicated nanophotonic structures, we have seen many new approaches to optical devices, including photonic crystals [1], metamaterials [2 – 4], nanometallics and plasmonics [5], the merging of ideas, such as antennas [6, 7] and waveguides [7 – 11] from radio-frequency devices into optics, and devices designed with fully arbitrary approaches to perform specific functions on multiple beams or wavelengths [12 – 15]. We are also seeing new demands for compact and novel optical functions such as mode converters for free-space communications [16, 17] or for coupling to multi-spatial mode optical fibers [18 – 21]; very compact wavelength splitters for optical interconnects [12, 13, 22], wavelength networks [23, 24] or spectroscopy; novel kinds of optical isolators [25]; and devices generally operating at deeply sub-wavelength scales [6 – 11, 26, 27].

How we exploit these new technological opportunities to address these many applications is a challenging design problem, especially for devices that must controllably map multiple inputs to multiple outputs. Here, to help address such design challenges, we consider one key question: How complicated does the optical component have to be? We need the answer for two reasons: (i) we want optical devices only to be as complicated to fabricate as necessary, and (ii) we want to make the device as easy to design and simulate as possible. In previous work on how much material a given device requires [28, 29], we considered some specific cases of complexity of the number of modes that needed to be controlled. In this paper, we examine generally the complexity required for linear optical components. Note we are here only considering the complexity that a design needs to have; the issue of how arbitrary linear components could be designed or configured will be considered elsewhere [30, 31].

First, in Section 2, we discuss complexity of an object in general. In Section 3, we discuss the mathematics of linear optical components generally. Here, we build on a formalism in which any linear optical device can be written as a mode converter [32]. In Section 4, we proceed to categorize optical devices by the complexity that they need and that different kinds of approaches can support, and give examples of complexity in various categories, including mode converters and various frequency demultiplexers. In Section 5, we draw conclusions.

## 2. Design complexity

When asking how complicated a device must be to perform some function, we encounter a very basic question: Can we measure or count how complicated something is? We could ask this in two different flavors: How complicated was it to make the object, and how complicated was it to design the object?

In both cases, there is arguably no definite answer because we do not know how far back to go in counting complexity. In fabrication, do we just count the activities required to make the object given the starting materials, or do we count also the steps required to make the materials themselves, and so on in a possibly endless regress including purifying starting chemicals, and extracting raw materials from the ground, and so on? In design, we have a similar problem. One extreme strategy might be to design the object atom by atom. This strategy could take a very long time. Another extreme strategy is just to keep looking until we find something that does the function we need. We could search through our optics for a 3.9 cm focal length lens. We might find one with the first object we check, or we might keep looking in vain. There is in general no way of quantifying in advance how many steps it takes to find something. Of course, we might regard merely finding something as being a weak form of design, but trying various things we find about us, possibly in combinations, is a time-honored design approach.

What we can quantify, however, is, given some starting set or repertoire of objects, how complicated it is to design any one out of a category of final objects. For example, given a repertoire of toy bricks, how complicated is it to design any member of a finite category of final toy houses, such as houses made with only so many bricks, or houses made according to some architectural design rules that constrain designs in some definite way? Such a question is generally answerable.

The example with toy bricks is one where we could possibly count a number of binary or discrete decisions that have to be made. In optics, we may instead be choosing analog values – real numbers, such as positions, lengths, or refractive indices – and we want to quantify how many of these we need to choose if we are to design any one of a category of devices. Below, therefore, we establish what we call device "complexity numbers" $N_D$ – the number of real numbers we need to specify to allow us to design any device within a given category.

In what follows, we attempt both to define some useful categories of optical devices and quantify the associated complexity number $N_D$, and we examine the usable complexity of various classes of devices. Before leaving this general discussion, we introduce two other concepts.

*2.1 Inherent functionality*

Inherent functionality is functionality or capability that is possessed by the object or class of objects that we do not explicitly design into it or them. When we choose initial components from some set, we have not designed the inherent capabilities of those starting components – we are merely exploiting them. In optics, we may choose to start with "blocks" like simple lenses, gratings and mirrors. These objects may have substantial useful inherent functionality. These specific ones have the general property that, once we choose in design what the component does for one beam (for example, by choosing its focal length, we design a lens so that it focuses an on-axis beam onto a particular back focal plane), then that designed component also happens to do similarly useful things for other input beams (for example, now focusing input beams at different angles onto different parts of the output plane). What the lens does for those other beams is an inherent property or functionality of the lens. Note that, given that we chose to use a lens to address our problem of interest for one beam, we do not get to choose with any substantial degree of independence what is to happen to the other beams. Similarly, once we choose the angle at which one beam bounces off a mirror (by choosing the mirror angle), we have defined what happens to input beams of other angles.

In many cases, the inherent functionality is not what we want; for example, a broad range of starting devices considered for slow light, including atomic vapors, optical resonators or sets of identical resonators, and photonic crystals, have the desirable functionality of delaying a pulse through group delay, but have the undesirable inherent functionality of distorting pulses [33]. A similar problem occurs in the closely related [34] superprism devices (see, e.g., [12, 13, 34 – 37]), which rely on group delay to shift a beam. Those made from periodic structures (e.g., [35 – 37]) generally have the undesirable inherent functionality of distorting the beam shape. Such problems can be alleviated by going to custom-designed non-periodic structures that avoid relying on the (here, undesirable) inherent functionality of periodic or simply resonant device structures and that exploit a larger number of designed degrees of freedom [12, 13, 34].

Once we have finished designing the device, the resulting complete device will still have some response for conditions other than those for which it was designed, and that response is inherent functionality of the final device. That inherent functionality, too, may be useful (e.g., similar response for wavelengths other than the design wavelength) or undesirable (e.g., chromatic aberration).

As we discussed above, we cannot generally quantify inherent functionality. And, we should acknowledge that much of the art of design is in choosing good starting objects with useful inherent functionalities or that lead to finished devices with other useful inherent functionalities, a process whose complexity again cannot be uniquely quantified.

*2.2 Externalizing functionality*

Sometimes when we design a system such as an optical one, we will decide that some of the functionality is best pushed out of the optical system into some other system. For example, we might be making some wavelength separator to route different wavelength channels to different outputs. The optical system might be a simple grating that puts short wavelengths into channels on the left, over to long wavelengths on channels to the right. First, because the outputs are evenly spaced with frequency, we have to arrange a matching spacing of detector or waveguide positions, which is pushing functionality into choices we have to make in the mechanical design. More substantially, we might have wanted the channels ordered in a different way, with one specific channel going to the fiber for Chicago and another going to the New York fiber, which might not be the order from the optical device. Then we need to follow the wavelength separating grating with some fiber patch panel or some additional set of optical or electrical switches to accomplish the routing we actually want. In that case, we could say that we have externalized part of the functionality of the system, pushing it outside the specific system (say, the grating) that we are designing.

This is a common design phenomenon, of course, but here we need to recognize at least when we are doing it. One recent example of externalizing functionality is a tunable detector whose tunability comes from the way that we electrically add signals from a multiple element detector in an interference pattern; the optics is fixed, but the device is rapidly tunable and programmable using electronics [23, 24]. Another recent example is a multiwavelength communications channel where individual optical filters are tuned over only a small section of the overall bandwidth of interest because of power dissipation limitations; the final signals are sorted to their correct actual destinations using an electronic circuit to perform bit re-shuffling [38]. Coherent communications externalizes functionality to digital signal processing [19], in part because we do not know how to design or fabricate some linear optical devices.

## 3. Mathematical preliminaries

To address complexity in optical devices, we start by writing linear optical devices in a general mathematical form based on the singular value decomposition (SVD) of the linear operator that describes the device [32]. Then, by counting the numbers required to specify the SVD, we quantify how complicated some categories of optical devices are. Some categories cannot solve particular problems because they cannot embody enough complexity in design, and others categories may be much more complicated than we need.

The fields, waves and devices we consider can vary in space, time, wavelength or frequency, and possibly other attributes, such as polarization or spin. The formalism is very general, and can be applied to any linear wave problem, including acoustic, electromagnetic, or even quantum mechanical waves. For definiteness and ease of visualization, we mostly discuss simple monochromatic spatial examples and/or spectral filters and demultiplexers.

Previously [32], we showed how any linear optical device can be described by a linear device operator $\mathsf{D}$ that takes an input function $|\phi_I\rangle$ and generates a corresponding output function $|\phi_O\rangle$,

$$|\phi_O\rangle = \mathsf{D}|\phi_I\rangle \qquad (1)$$

We can essentially always perform the SVD of $\mathsf{D}$ [32] to yield an expression

$$\mathsf{D} = \sum_m s_{Dm} |\phi_{DOm}\rangle\langle\phi_{DIm}| \qquad (2)$$

or, equivalently,

$$\mathsf{D} = \mathsf{V}\mathsf{D}_{diag}\mathsf{U}^\dagger \qquad (3)$$

where $\mathsf{U}$ ($\mathsf{V}$) is a unitary operator that in matrix form has the vectors $|\phi_{DIm}\rangle$ ($|\phi_{DOm}\rangle$) as its columns and $\mathsf{D}_{diag}$ is a diagonal matrix with complex elements (the singular values) $s_{Dm}$. (We use Dirac notation here for the linear algebra. See, e.g., Refs. [32] and [39]). The sets of functions $|\phi_{DIm}\rangle$ and $|\phi_{DOm}\rangle$ corresponding with non-zero singular values $s_{Dm}$ each form orthogonal sets that are complete in the input space $H_I$ and output space $H_S$ respectively [32].

## 4. Device complexity

To evaluate complexity meaningfully, we need to establish the number of basis functions or modes at the input and at the output, that is, the dimensionalities $M_I$ and $M_O$, respectively, of the input and output spaces $H_I$ and $H_O$. [In what follows, we use the terms "modes" and "basis functions" interchangeably (see, e.g., Ref. [39], pp. 516-

518)]. We may already know that there are $M_I$ input orthogonal input modes that can couple into the device and $M_O$ orthogonal output modes than can couple out of it. For example, we might be considering a monochromatic spatial mode problem with waveguides in and out of the device; the input and output waveguides might only support $M_I$ and $M_O$ spatial modes respectively (Fig. 1). If $M_I$ and $M_O$ are not initially obviously well defined, we can calculate them in any given situation, with some assumptions (see Appendix). Henceforth, we presume we know $M_I$ and $M_O$.

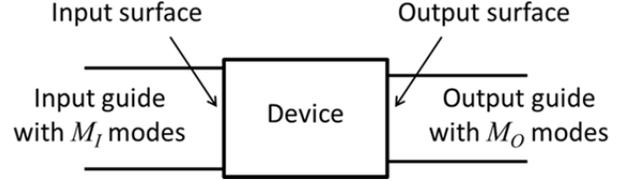

Fig. 1. Sketch of an example device with an input waveguide with $M_I$ modes and an output waveguide with $M_O$ modes. Here the input (output) mathematical space corresponds to the functions on the input (output) surface.

The question here is what is the number of mathematical parameters (i.e., real numbers) we must choose to specify any device within some device category? We start by considering the most general possible device that operates on $M_I$ orthogonal input modes and gives outputs into $M_O$ orthogonal output modes. Any such linear optical device is obviously completely describable by the $M_O(\text{rows}) \times M_I(\text{columns})$ matrix $\mathsf{D}$ of (complex) numbers that gives the $M_O$-dimensional output vector $|\phi_O\rangle$ in response to the $M_I$-dimensional input vector $|\phi_I\rangle$ as in Eq. (1). Therefore, $M_O M_I$ complex numbers, or the "complexity number"

$$N_D = 2M_O M_I \qquad (4)$$

of real numbers, are sufficient to specify the device.

Though Eq. (4) is correct for the most general possible linear device, it is an over-estimate for many useful categories for three reasons. (i) The devices in the category we want to make may be simpler than the most complicated device describable by such a matrix, so we could construct the matrix using fewer independent real parameters. (ii) The way we make the device may not allow us usefully to specify sufficient parameters to make such an arbitrary device. Volume holograms [40, 41], some spectral filters [12, 13, 34, 42], and recent design exercises in mode converters [14] are possible examples of devices that could approach the full complexity offered in Eq. (4), but, as we will see below, many optical devices, such as lenses, mirrors, gratings and thin holograms, generally are not capable of offering the level of complexity suggested by Eq. (4). (iii) The inherent functionality of the starting components may make the problem much easier in practice (though often this is because we are also externalizing functionality). We now clarify this discussion and quantify $N_D$ by examining the SVD, Eq. (2) or (3), of the device matrix.

### 4.1 Maximally connected and maximally functional devices and mode coupling number

First, we define two new terms – maximally connected devices and maximally functional devices – and a related concept, mode coupling number.

*Maximally connected devices and mode coupling number*

Performing the SVD of the $M_O \times M_I$ device matrix $\mathsf{D}$, in general we get a number $M_{min}$ of singular values that is the smaller of $M_O$ or $M_I$, that is

$$M_{min} = \min(M_O, M_I) \qquad (5)$$

because this is the number of diagonal elements in a rectangular $M_O \times M_I$ matrix.

If all of these $M_{min}$ singular values are non-zero, then we call the device "maximally connected". That means that the device does possess finite coupling from the input to the output for the largest possible number of orthogonal input to output connections allowed by the numbers $M_I$ and $M_O$. As we will see explicitly, not all devices are maximally connected.

We can usefully define a "mode coupling number", $M_C$, which is the number of non-zero singular values of the device operator $\mathsf{D}$, and which counts the number of orthogonal input to output connections. A device is maximally connected if $M_C = M_{min}$. If $M_C < M_{min}$, the device is not maximally connected.

*Maximally functional device*

A maximally functional device is one for which we can choose enough parameters of the right kind physically and in the right places in the mathematics to allow any linear mapping between input and output for the $M_C$ orthogonal input and output connections that the device supports. As we will see explicitly, not all linear optical devices are maximally functional (in fact, most are not).

As we will show by examples, a device can be maximally functional even if it is not maximally connected, and *vice versa*, so these are independent concepts.

### 4.2 Maximally connected, maximally functional device

For the sake of definiteness in discussion, we presume for the moment that the number of input modes $M_I$ exceeds the number $M_O$ of output modes (we can construct similar arguments in the opposite case if we wish) and that this device is maximally connected, which in this case means $M_C = M_O$. Performing the SVD of $\mathsf{D}$, we could write the resulting matrices in two ways (Fig. 2).

One way would be to write $\mathsf{U}$ as an $M_I \times M_I$ matrix and $\mathsf{D}_{diag}$ as a $M_O \times M_I$ matrix, with the $M_O$ singular values down the diagonal and zeros everywhere else in this matrix. Necessarily, the right-most $M_I - M_O$ columns of this matrix $\mathsf{D}_{diag}$ are zero. Hence, in the matrix product $\mathsf{D}_{diag}\mathsf{U}^\dagger$, the lowest $M_I - M_O$ rows of $\mathsf{U}^\dagger$ never have any influence on the resulting matrix product. Consequently, it is simpler, and equivalent, to write $\mathsf{D}_{diag}$ as a $M_O \times M_O$ square matrix (or, more generally, an $M_{min} \times M_{min}$ matrix), and to write $\mathsf{U}$ as an $M_I \times M_O$ matrix (making $\mathsf{U}^\dagger$ a $M_O \times M_I$ matrix). The $M_I - M_O$ discarded columns of $\mathsf{U}$ correspond to input functions such that no linear combination of them every leads to any output from the device. (Technically, this new form makes $\mathsf{U}$ not strictly a unitary matrix; $\mathsf{U}^\dagger\mathsf{U}$ is an identity matrix (of dimensions $M_O \times M_O$), but the $M_I \times M_I$ matrix $\mathsf{U}\mathsf{U}^\dagger$ is not in general unitary, though this causes no formal problems here.) Writing the matrices in this second way makes it easier to evaluate the complexity number $N_D$; we avoid counting numbers that do not influence the device behavior.

(a)
$$\mathsf{D} = \begin{bmatrix} v_{11} & v_{12} \\ v_{21} & v_{22} \end{bmatrix} \begin{bmatrix} s_{D1} & 0 & 0 & 0 \\ 0 & s_{D2} & 0 & 0 \end{bmatrix} \begin{bmatrix} u_{11}^* & u_{21}^* & u_{31}^* & u_{41}^* \\ u_{12}^* & u_{22}^* & u_{32}^* & u_{42}^* \\ u_{13}^* & u_{23}^* & u_{33}^* & u_{43}^* \\ u_{14}^* & u_{24}^* & u_{34}^* & u_{44}^* \end{bmatrix}$$

(b)
$$\mathsf{D} = \begin{bmatrix} v_{11} & v_{12} \\ v_{21} & v_{22} \end{bmatrix} \begin{bmatrix} s_{D1} & 0 \\ 0 & s_{D2} \end{bmatrix} \begin{bmatrix} u_{11}^* & u_{21}^* & u_{31}^* & u_{41}^* \\ u_{12}^* & u_{22}^* & u_{32}^* & u_{42}^* \end{bmatrix}$$

Fig. 2. Illustration for the case of $M_I = 4$ and $M_O = 2$ of two ways of writing the matrix product $\mathsf{D} = \mathsf{V}\mathsf{D}_{diag}\mathsf{U}^\dagger$. Both methods (a) and (b) lead to the same results for any multiplication of the matrix $\mathsf{D}$ by a vector. Method (a) has strictly unitary square matrices for both $\mathsf{U}$ and $\mathsf{V}$, but it has more parameters than the reduced version (b). Note that the bottom two rows of the right-most matrix in (a) never enter into any calculation in multiplying a vector by $\mathsf{D}$.

The columns of $\mathsf{U}$ are normalized orthogonal vectors. Obviously, to specify an arbitrary vector in an $M_I$-dimensional space requires $M_I$ complex numbers ($2M_I$ real numbers). The normalization of these vectors fixes one real amplitude coefficient, leaving $2M_I - 1$ free real numbers. The absolute phase of any of the columns in $\mathsf{U}$ (or $\mathsf{V}$) is also of no importance. (As usual, the overall phase of the eigenfunctions is arbitrary in the eigenvalue problems we solve to construct the SVD.) The relative phase of the output $|\phi_{DOm}\rangle$ for a given input $|\phi_{DIm}\rangle$ can be set by choosing the phase of the corresponding singular value $s_{Dm}$, which we are free to do mathematically. Hence, without loss of generality, we can set the overall phase of each of the columns in $\mathsf{U}$ (and $\mathsf{V}$); for example, we could choose the first non-zero element in each column to be real.

Hence we are left with $2M_I - 2$ real coefficients to specify one column in $\mathsf{U}$. The first column we choose requires this many real numbers to specify it. The second column we choose also has to be orthogonal to the first, which requires that both the real and imaginary parts of the inner product (e.g., overlap integral) between the first and second columns are zero, thus reducing the number of (real) free parameters for the second column function by 2, to $2M_I - 4$. The third column similarly has to be orthogonal to both the first and the second chosen columns in a given matrix, reducing its free real parameters to $2M_I - 6$, and so on, with the $q$-th chosen column having $2M_I - 2q$ free real parameters. Counting for all the $M_O$ columns of $\mathsf{U}$ gives a total of

$$N_\mathsf{U} = \sum_{q=1}^{M_O} 2M_I - 2q = 2M_O M_I - M_O(M_O + 1) \quad (6)$$

real numbers.

For the (square) $M_O \times M_O$ matrix $\mathsf{V}$, we count similarly, adding $2M_O - 2$, $2M_O - 4$, and so one, in this case all the way to $2M_O - 2M_O = 0$ free real numbers in the final ($M_O$-th) column, for a total of

$$N_\mathsf{V} = \sum_{q=1}^{M_O} 2M_O - 2q = M_O^2 - M_O \quad (7)$$

real numbers to specify the matrix $\mathsf{V}$. The matrix $\mathsf{D}_{diag}$ requires $2M_O$ real numbers to specify its diagonal complex (singular value) elements, so the total number of real numbers required to specify the matrices $\mathsf{U}$, $\mathsf{V}$, and $\mathsf{D}_{diag}$ is a complexity number

$$\begin{aligned}N_D &= 2M_O M_I - M_O(M_O + 1) + M_O^2 - M_O + 2M_O \\ &= 2M_O M_I\end{aligned} \quad (8)$$

which is exactly what we would expect for the construction of an arbitrary $M_O \times M_I$ matrix of complex numbers, as in Eq. (4). Constructing a similar argument for the case where the number of output modes $M_O$ exceeds the number of input modes $M_I$ leads to the same result of the bottom line in Eq. (8). In this case, $\mathsf{D}_{diag}$ and $\mathsf{U}$ are both $M_I \times M_I$ matrices, and we write $\mathsf{V}$ as an $M_O \times M_I$ matrix.

So far, then, the counting using the SVD form of the matrix $\mathsf{D}$ yields the same results as the obvious counting of the number of real numbers required to specify an arbitrary $M_O \times M_I$ matrix. Since we have enough parameters here in a suitable mathematical form to specify an arbitrary $M_O \times M_I$ matrix, and hence we can define an arbitrary linear mapping for all $M_C = M_{min}$ orthogonal input-to-output channels, this approach is describing a maximally functional device category. Because $M_C = M_{min}$, this approach is also describing a maximally connected device category.

Where the SVD form becomes more obviously useful in the counting is when we consider either device categories that are not maximally connected or that are not maximally functional or both.

### 4.3 Sub-maximally connected device example – single-mode converter

Suppose we want a device that takes one specific (normalized) input beam $|\phi_{DI1}\rangle$ in the $M_I$-dimensional input space and generates as a result one specific output beam proportional to the (normalized) function $|\phi_{DO1}\rangle$ in the $M_O$-dimensional output space with some amplitude $s$, i.e., an output $s|\phi_{DO1}\rangle$, and that for any other orthogonal input beam, the device generates no output. This describes an ideal mode converter, mode coupler or spatial filter for converting only one mode to another specific mode.

In this case, the SVD can be written directly. The elements of the $M_I \times 1$ matrix $\mathsf{U}$ are the elements of vector $|\phi_{DI1}\rangle$. Similarly, the elements of the $M_O \times 1$ matrix $\mathsf{V}$ give the vector $|\phi_{DO1}\rangle$. The matrix $\mathsf{D}_{diag}$ is the $1 \times 1$ matrix containing the sole singular value $s_{D1} = s$. Mathematically, we are constructing an $M_O \times M_I$ matrix by taking the (outer) product $|\phi_{DO1}\rangle\langle\phi_{DI1}|$ of these vectors, multiplied by the complex number $s$. If we have input and output spaces that are both multimoded – i.e., $M_I > 1$ and $M_O > 1$ – this device is not maximally connected (or, equivalently, it is sub-maximally connected) because the number of non-zero singular values $M_C$ ($=1$) $< M_{min}$; equivalently, $M_C < M_I$ and $M_C < M_O$. The device category is, however, maximally functional in that it has enough parameters specifying it to allow the design of any such device with mode-coupling number $M_C = 1$ for arbitrary input and output beams in these spaces.

As before, because of normalization and fixing the phase of the (sole) column in each of the matrices $\mathsf{U}$ and $\mathsf{V}$, we require $2M_I - 2$ and $2M_O - 2$ real numbers respectively to specify these matrices, and $2$ real numbers to specify the (sole) singular value that makes up the matrix $\mathsf{D}_{diag}$. Hence, altogether, we need a complexity number

$$N_D = 2(M_O + M_I - 1) \quad (9)$$

of real numbers to specify the operator $\mathsf{D}$ for a device that is to perform any specific function in this category. Note that this number of parameters is *not* in general the $2M_O M_I$ or Eq. (4). The fact that $N_D < 2M_O M_I$ (when the input and output spaces are multimoded) reflects the fact that this device is sub-maximally connected. Here there are possible input functions in the input space that lead to no output. We illustrate an approximate implementation of such a device in section 4.7 below.

### 4.4 General maximally functional device

From the examples above, it is straightforward to see how to construct the matrices for a device category that converts from multiple arbitrarily chosen input modes to multiple corresponding arbitrarily chosen output modes. We presume input and output spaces with $M_I$ and $M_O$ dimensions respectively, and we want a device category that makes $M_C$ arbitrary orthogonal connections between the two, where the device is not necessarily maximally connected (i.e., $M_C$ may be less than both $M_I$ and $M_O$). The matrices $\mathsf{U}$ and $\mathsf{V}$ are $M_I \times M_C$ and $M_O \times M_C$ dimensional, respectively, and $\mathsf{D}_{diag}$ is an $M_C \times M_C$ diagonal matrix. The columns of $\mathsf{U}$ ($\mathsf{V}$) are normalized versions of the orthogonal input (output) modes of interest with fixed overall phases according to some rule we choose, and we have $M_C$ chosen complex connection amplitudes as represented by the (singular value) diagonal elements in $D_{diag}$.

As we established in Eq. (6) for the $M_I \times M_O$ matrix $\mathsf{U}$ in the case of a maximally connected maximally functional device above, the number of real numbers required to specify the $M_I \times M_C$ matrix $\mathsf{U}$ here is similarly $2M_C M_I - M_C(M_C + 1)$. For the $M_O \times M_C$ matrix $\mathsf{V}$ we require a further $2M_C M_O - M_C(M_C + 1)$ real numbers. Adding the $2M_C$ real numbers required to specify the singular values in $D_{diag}$ gives a total complexity number of

$$N_D = 2M_C(M_I + M_O - M_C) \qquad (10)$$

for this category. This expression covers both the maximally and sub-maximally connected cases, giving the same answers as Eqs. (8) and (9) in those specific cases.

### 4.5 Neglecting output phase for a maximally functional device

Many optical components do not care about the relative phase of different orthogonal outputs. For example, when separating multiple beams to different detectors to measure power, the phase of the beams hitting the detectors is irrelevant. Similarly, in a wavelength demultiplexer, we typically do not care about the phases of the separated wavelengths going into different output channels. In such cases, we therefore do not need also to specify the phase of the singular values, so we can reduce the required number of degrees of freedom by $M_C$, leading to a complexity number

$$N_D = 2M_C\left(M_I + M_O - M_C - \frac{1}{2}\right) \qquad (11)$$

instead of Eq. (10). For the single-mode converter above, if the phase of the output is unimportant, then we similarly subtract 1 from Eq. (9), in agreement with Eq. (11).

### 4.6 Multiple spatial mode converter example

As an example, suppose we wanted the device to be able to take two different complicated modes, such as two arbitrarily chosen Gauss-Laguerre beams (orbital angular momentum beams) from some set, and explicitly turn each one into a different spot on an output plane. The device input space would have to be large enough to be able to distinguish each of these beams from other beams in the set. If we needed to distinguish each beam from, say, 19 other orthogonal beams, then, the input space would have to have at least a dimensionality $M_I$ of 20. On the device output, we can have a dimensionality $M_O = 2$ because we only want two different orthogonal couplings, and we can ideally choose mode coupling number $M_C = M_O = 2$. To make the device maximally functional, so we can choose any two orthogonal functions in the 20 dimensional input space (including orthogonal linear combinations of the modes from the input space) and put them where we want in the output space, we use Eq. (10) to calculate the complexity number $N_D = 2M_C(M_I + M_O - M_C) = 2M_I M_O = 80$ if we care about the phase of the outputs relative to the inputs. If we only care about getting the power to go into the required output modes, then we can use Eq. (11), which gives $N_D = 78$. These are then the numbers of parameters we need to specify in the device if we are going to be able to make arbitrary choices of two orthogonal input functions in the 20 dimensional input space and create two arbitrarily chosen orthogonal outputs.

There are relatively few clear examples of actual attempts at spatial mode converters that convert from multiple arbitrary input modes to multiple arbitrary output modes, and where one can count the number of degrees of freedom used in design. One such example is the design of a device to convert, in a 2-dimensional photonic-crystal-like device, from the three different modes of a three-moded input guide to three different single-mode output guides [14]. In our terminology, this device is certainly maximally connected (connecting all three input modes to the output space). Though this work did not check that the output guides could be arbitrarily positioned and that arbitrary assignment of input modes to output positions would be possible, we might reasonably conjecture these capabilities in this structure because there is no obvious inherent functionality that constrains the device otherwise. With that conjecture, from the specific structure of this device [14] we could reasonably expect that there are 15 total positions to put single-moded output guides on the right hand side of this structure. Hence we conjecture this device design approach could represent one that is maximally functional and maximally connected, with $M_C = M_I = 3$ and $M_O = 15$. Using Eq. (11), we calculate we need to specify $2M_I M_O - M_C = 87$ real numbers in design to allow us full design freedom if the relative phases between inputs and outputs are unimportant.

This particular device was designed using 105 binary variables, each representing the presence or absence of a particular pillar on a 7x15 grid of positions. Though it is not obvious how to compare binary and continuous real degrees of freedom, 105 is at least in the same overall magnitude as 87, being somewhat larger as we would expect in using binary rather than real numbers. We note, though, that it is merely conjecture that this particular design approach is fully functional for mapping into 15 different output modes.

### 4.7 Maximally connected devices with sub-maximal functionality – "mask-based" devices

So far, we have discussed only maximally functional device categories, where we connect a given number of orthogonal input functions to orthogonal output functions with arbitrary choice of the functions in the input and output spaces in each case. Now we look at a particularly important example category of sub-maximally functional devices.

A broad and important class of the optical devices that we use, such as (thin) lenses, gratings, transparencies (i.e., objects we are projecting) and thin holograms, can be thought of, at least approximately, as devices where we multiply the input field (e.g., $\phi_I(x,y)$) at a given position $(x,y)$ on the device input surface by a position-dependent complex factor (e.g., $D(x,y)$) to get the output field (e.g., $\phi_O(x,y)$) at the corresponding point on the device output surface. For example, a two-dimensional device with scalar waves might obey the relation

$$\phi_O(x,y) = D(x,y)\phi_I(x,y) \qquad (12)$$

We could call such devices collectively "mask-based" devices because their behavior is defined by a single "mask" function, e.g., $D(x,y)$.

In the time domain, if a modulator in a single-mode fiber had a specified transmission "mask" function $D(t)$ of time $t$, then we could similarly have a relation between the input $\phi_I(t)$ and the output $\phi_O(t)$ of the form

$$\phi_O(t) = D(t)\phi_I(t) \qquad (13)$$

(This example corresponds to a device made with non-dispersive materials so there is no temporal memory in the $D$ function.)

Note that, though relations such as Eqs. (12) and (13) are linear, they are by far not the most general linear relation we could have between input and output functions. For scalar functions of one continuous variable, most generally we would instead have

$$\phi_O(t) = \int G(t,t_1)\phi_I(t_1)\,dt_1 \qquad (14)$$

(We could construct a similar general relation for the two-variable case of Eq. (12).)

Writing Eq. (14) in matrix-vector terms would give

$$|\phi_O\rangle = \mathsf{G}|\phi_I\rangle \qquad (15)$$

If we think of the vectors $|\phi_I\rangle$ and $|\phi_O\rangle$ as each being columns of numbers giving the values of these functions at successive (closely spaced) values of the variable $t$ (or the variables $x$ and $y$), then the matrix $\mathsf{G}$ is fully populated with elements that in general are non-zero. In that case, an input field at one specific point can in general lead to finite output fields at all output points. Mask-based devices do not behave this way, however, because the response within the device is local – an input at a given position or time gives an output only at the same position or time.

*Counting parameters for mask-based devices*

We can view mask-based devices in such a "position" basis; that is, we work with basis functions that are essentially delta functions or strongly localized functions of position $(x,y)$ for devices described by Eq. (12) or time $t$ for devices described by Eq. (13). Then, in the matrix form of relations like Eqs. (12) and (13) (i.e., Eq. (1)), the matrix $\mathsf{D}$ is diagonal; the off-diagonal elements are all zero, reflecting the fact that such a local "mask" operation does not generate output fields at other points on the mask. We should therefore expect that such mask-based devices have a restricted functionality – not all conceivable linear relations between input and output are possible for such devices and so they have sub-maximal functionality. Because the operator $\mathsf{D}$ is diagonal in such localized basis sets, we have already effectively performed the SVD of $\mathsf{D}$.

For actual optical situations, the input and/or output spaces might not support very strongly localized functions. For example, if we had a multimode waveguide for the input that supported, say, 4 modes, we typically would not be able to use linear combinations of those to make a function that was localized to anything smaller than ~ ¼ of the waveguide cross-section area. Nonetheless, we can still conjecture that, at least for input and output spaces that have relatively large numbers of basis functions, we can use those basis functions to form functions ("spots") that are small enough that we can use our approach here, at least approximately and conceptually, to count the number of degrees of freedom available to us in mask-based devices.

Presuming we have equal numbers of input and output functions, i.e., $M_I = M_O = M$, then we conjecture we can use these basis functions to form $M$ different functions $|\phi_{DIm}\rangle$ ("spots") in the input space and similarly $M$ different functions $|\phi_{DOm}\rangle$ ("spots") in the output space that (i) are strongly localized near specific points, $\mathbf{r}_m$, in space (the same point $\mathbf{r}_m$ for both functions $|\phi_{DIm}\rangle$ and $|\phi_{DOm}\rangle$ for a given $m$), and (ii) are approximately orthogonal to one another in a given space (at least because that do not overlap strongly, being localized round sufficiently different points in space).

Now we conjecture also that we can form the singular values $s_{Dm}$ associated with these "spots" $|\phi_{DIm}\rangle$ and $|\phi_{DOm}\rangle$ near points $\mathbf{r}_m$ by some effective averaging of the function $D(x,y)$ over these spots around each point $\mathbf{r}_m$. This argument is somewhat approximate and conjectural; the only conclusion we want to draw here, however, is that, no matter how complicated the function $D(x,y)$ is, it only defines ~ $M$ different (complex) values, here represented by the ~ $M$ singular values $s_{Dm}$ that go into controlling the final behavior of the device. Note in particular that our argument gives an answer

$$N_D \simeq 2M \qquad (16)$$

for the complexity number of real parameters to specify the device, not the $2M^2$ of the maximally functional device as in Eq. (8).

One might ask why we are not also counting the number of real numbers to define the columns of the matrices $\mathsf{U}$ and $\mathsf{V}$ in the present counting. The answer is that the locality of the response of mask-based devices already enforces the form of those columns (and equivalently of the form of the $|\phi_{DIm}\rangle$ and the $|\phi_{DOm}\rangle$); by our conjectures, those columns already necessarily represent the maximally localized orthogonal functions that can be generated from the basis sets of the input and output spaces, and there is essentially only one way of forming those sets. [One technical exception to this rule is if we have multiple different positions with the same singular value (i.e., the same transmission or reflection); then we are free to choose linear combinations of these positions in making up the basis functions. However, such a case is relatively restrictive since it corresponds only to uniform transmission over such a set of positions. We could also permute the columns of the matrices U and V (by the same permutation) and similarly permute the order of the singular values in $D_{diag}$, but such a permutation merely corresponds to a different labeling of the same physical spots, not to any change in the device, so such permutations are arbitrary and need not be considered.]

*Limitations of mask-based devices*

We do not attempt to answer the general question of what functions mask-based devices can implement, but we can make some observations. Suppose first we consider a general mask based device with $M$-dimensional input and output spaces, and suppose we want to run it at what would be its "diffraction" or "aperture filling" limit for a spatial device – that is, we illuminate it with a fully $M$-dimensional input function, one that in general has non-zero amplitudes for all the $M$ (localized) input SVD functions $|\phi_{Dlm}\rangle$. Then, to specify the desired output for this one beam requires that we specify all the $\sim 2M$ real numbers of Eq. (16). So, in defining what this device is to do for one "diffraction limited" or aperture-filling input, we have completely specified the device; there are no degrees of freedom left to define what the device does for any other inputs. To emphasize, we now have no ability to choose what happens to any other input function.

This result might seem surprising and even counter-intuitive. For a device such as a lens, for example, we obviously can use it productively for a large range of different inputs, each of which may fill the lens aperture. Suppose, though, we think of a lens as a device that we design so that, when illuminated by a plane wave propagating along the optical axis (i.e., in a direction perpendicular to the lens plane), it generates a single diffraction-limited spot in the center (i.e., along the optical axis) of a plane at a distance $f$ behind the lens (the focal plane). Of course, we know how to do this; the lens is designed to impose a phase delay that varies quadratically with distance from the center of the lens, with focal length $f$. But note that we have automatically set what the lens does with a plane wave at any other incident angle. At least in a paraxial approximation, we simply form other similar spots in the focal plane, displaced angularly by (minus) the angle of the input beam to the optical axis; our design of what happens for one aperture-filling beam on a simple lens has defined what happens for all other beams. (Note any particular input beam could be made up out of a linear combination of plane waves at different angles, so by this one design for one plane wave, we have defined the optical device for all conceivable inputs.)

What happens for other beams is a consequence of the inherent functionality of a lens device. Though we may have been wise enough to choose a lens as the component to use in the system, we cannot quantify its inherent functionality, as discussed above, nor (at least for a single thin lens) can we control its effects on other beams once we have designed what happens for one beam.

We mentioned above similar inherent functionality for another "mask-based" device – namely, a plane mirror. A simple plane mirror takes an input beam at one specific angle and converts it to a beam at another specific angle (as given by specular reflection). Once we have chosen this angular change for any one input beam (by choosing the physical angle of the mirror), we have defined the angle we get for any other angle of input beam. A (reflective) grating shows similar behavior to a mirror except that the angle (or angles) of the output beam(s) for a given frequency of input is (are) not necessarily the specular reflection angle. Once we have chosen the behavior for one input beam, we have essentially defined also what happens for other angles of input beam of the same frequency. We have also generally determined the output angles for beams of different wavelengths.

A general mask-based device, such as we might implement with a thin hologram or diffractive optical element, can show a variety of different behaviors, but has the same limitation. For example, it is straightforward to show that, at least in a standard Fourier-transform optical system that we might use with such an element [43], if we want outputs for different input beams to merely be shifted replicas of one another (e.g., we want identical spots at different positions when the element is presented with different input beams), then the input beams must simply be tilted versions of one another. This is an illustration that, if we define output beams we want for a mask-based device, once we have chosen what the input is that gives a specific one of those outputs, then the other inputs are all fixed. Equivalently, running the same mathematics backwards, if the inputs are a series of shifted versions of the same function (for example, identical spots at different positions), then the outputs are just the same beam at similar angular shifts.

We could argue that we could increase the size of the mask-based device so that we could get more degrees of freedom and hence so that we could define more and different mappings from the input space to the output space. For example, we could make a large mask-based device and divide it into $M$ sub-apertures each with $M$ usable elements for a total of $M \times M$ elements. We could illuminate each of those sub-apertures with a (necessarily) different beam, and we could separately control what the output would be for each of these sub-apertures. This approach is equivalent to making $M$ separate $M$-element mask-based devices. But as we make the overall device bigger, the SVD always retains the same form: It is diagonal with the local mask transmission factors as its singular values and with input and output matrices U and V whose columns are localized functions. We never get any real choice as to what the matrices U and V are. All of the free parameters in the design are taken up in specifying the singular values. Thus with a mask-based device alone we are never capable of designing a general optical $M$-mode element that corresponds to an arbitrary $M \times M$ matrix.

We can, of course, add further complexity by adding other optics to a system with mask-based devices to increase the functionality. For example, we can beam-split the input beam onto the *M* sub-apertures, perform *M* separate *M*-dimensional single-mode to single-mode conversions (as in the matched filter implementation below) and then combine the results into overlapping output beams using another beam splitter. This particular scheme, however, incurs two $1/M$ power loss factors in this emulation of a true maximally functional *M*-dimensional device with an $M \times M$-dimensional locally responding one.

Of course, the genius of the concepts of common mask-based devices such as lenses, gratings and mirrors, and of new concepts such as devices that unwrap orbital angular momentum beams [16] is that, despite this limitation of only being able to specify completely what happens to one input, these devices inherently perform useful functions for all *M* input modes. But, we do not have separate direct control over what happens to any beam other than the first (aperture-filling) one for which we choose to design.

*Matched filter implementation of single-mode converter*

As an illustration of a mask-based implementation of a device we have explicitly discussed, we can approximately implement the single-mode converter of section 4.3 above using a pair of masks (Fig. 3), working in the spirit of matched filters in Fourier optics (see Ref. [43], p. 248 et seq.).

We want to map a specific input mode $\phi_I(x,y)$ to a specific output mode. We construct input and output masks with (complex) transmission functions $\phi_{MI}(x,y)$ and $\phi_{MO}(x',y')$ respectively. If we choose $\phi_{MI}(x,y) \propto \phi_I^*(x,y)$, then the field to the right of the input mask is of the form $\phi_I(x,y)\phi_I^*(x,y)$, which we know is positive for all *x* and *y*. To the extent that a lens of focal length *f* performs an approximate Fourier transform into its back focal plane, then this positivity ensures we have a d.c. component in spatial frequency space, and hence a spot in the center of the aperture plane. Other input functions $\phi_{NI}(x,y)$ orthogonal to the desired input $\phi_I(x,y)$ lead to an output field $\phi_{NI}(x,y)\phi_I^*(x,y)$ from the mask that has no d.c. component because

$$\iint_{x,y} \phi_{NI}(x,y)\phi_I^*(x,y)\,dxdy = 0 \qquad (17)$$

for any such input $\phi_{NI}(x,y)$, and hence no spot in the center of the aperture plane. Hence, with a sufficiently small aperture, we can discriminate against all other orthogonal modes, leaving only one channel that is allowed to propagate through the aperture. The output field from the aperture has some form $\phi_A(x',y')$ after it passes through the output lens, a form that results from the spot shape in the aperture and the diffraction effects from the aperture. Provided that form $\phi_A(x',y')$ has no zeros in it over the output mask plane, which we can always arrange by making the aperture sufficiently small, then using an output mask of the form

$$\phi_{MO}(x',y') \propto \phi_O(x',y')/\phi_A(x',y') \qquad (18)$$

generates the desired output field $\propto \phi_O(x',y')$.

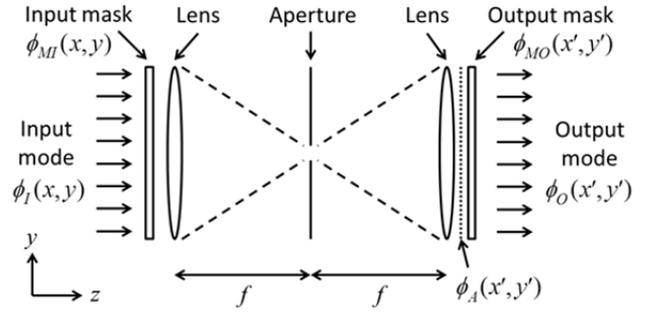

Fig. 3. Architecture for a single-mode converter based on Fourier optics. (The *x*-axis is into the plane of the drawing.)

Note that we have used $\sim 2M_I$ real numbers to specify the input mask sufficiently and $\sim 2M_O$ real numbers to specify the output mask sufficiently. We only need to specify absolute phase and amplitude on one or other mask, saving us formally two degrees of freedom, and hence giving us the answer of Eq. (9) for the complexity number $N_D$ as expected. Note that the aperture has made this device sub-maximally connected, approximately allowing only one channel to pass through the device and blocking other fields.

*Mask-based devices and phase*

In computer generated holograms to produce an output pattern, we typically want only to use a phase object to avoid absorbing power. If we only care about output intensity, however, we can allow the output phase to float, so we still have the *M* degrees of freedom in the mask phase to choose *M* intensities in an output pattern. Hence, consistent with the analysis here, we can make Dammann grating spot array generators [44], for example, and other useful phase-only holograms. We can also use such approaches to combine multiple laser gain media if the device is used within the overall cavity, because the phase of each individual laser gain medium output can float so as to give maximum power in the overall supermode [45].

*Frequency domain mask-based devices – fixed single-mode filters*

In a spectral filter, we may only be interested in one input spatial mode, such as a particular input beam or plane wave or a single mode in a fiber, and one output spatial mode, such as a transmitted or a reflected version of the input wave. If the filter is made from fixed materials – i.e., with no attribute of the filter changing in time – then the filter cannot create any new frequencies or transfer power from one frequency to another. As a result, the behavior of filter can be written in terms of some fixed transmission or reflection function $D(\omega)$ that depends only on frequency $\omega$. Quite generally, we can write the input in the input spatial mode as some function $\phi_I(\omega)$ of frequency, and similarly for the output $\phi_O(\omega)$. Then the relation between the input and the output can be written

$$\phi_O(\omega) = D(\omega)\phi_I(\omega) \qquad (19)$$

which is a "mask-based" relation that is local in frequency. Such a fixed filter therefore has the same counting of degrees of freedom as other mask-based devices and analogous limitations in the functions it can perform.

Of course, filters often do not have a physical function in their structure that is the direct analog of the mask functions $D(x, y)$ and $D(t)$ above; instead, they may be made from multiple dielectric layers, for example, with the function $D(\omega)$ arising as a result of a complicated set of reflections within the layers. However, any such filter can always be emulated in principle using a dispersive device such as a grating at the input, which separates the different wavelengths spatially, followed by a transmission mask that multiplies each different spatial point (and hence each different frequency $\omega$) by some number $D(\omega)$. Finally, we can run a similar dispersive device backwards to combine the different frequency components back into the one output spatial mode, thereby making the filter with an explicit physical mask of the form $D(\omega)$. (This is essentially the architecture commonly used [46] for shaping of short light pulses.)

As in the spatial case, this locality means that we do not have the freedom to choose the input and output functions in the SVD; those sets of orthogonal inputs that map to orthogonal outputs are frozen to be the single frequency functions. In general, therefore, using such a fixed filter we cannot in general make a device that can take multiple different chosen orthogonal spectra at the input, all filling the same bandwidth, and convert them to different chosen orthogonal spectra at the output. This conclusion would be obvious if the outputs were to contain different frequencies from the inputs, because we already know we cannot generate new frequencies with a fixed linear device. However, even if we never generate any new frequencies outside the bandwidth of interest, and even if we restrict the output at any given frequency to be of lower amplitude than the input at the same frequency for every spectrum of interest, we would face the same challenge, with no obvious way to accomplish this other than by power splitting to and from multiple different mask-based devices.

### 4.8 Wavelength demultiplexer examples

Wavelength demultiplexers offer a good example of non-trivial complexity calculation because of the constraints of the mask-based nature of optical filters in general (in the absence of wavelength conversion) and the dimensionality added through multiple spatial output channels.

Consider a device that runs with $M_I$ different wavelengths that all enter in one spatial mode, as in a single-mode optical fiber. Formally, the number of input spatial modes is then $M_{SI} = 1$. Different wavelengths of beam in one spatial mode are generally orthogonal functions in time. To be definite, consider a specific window for input times $t$, from time zero to time $t_I$, and consider waves of different (free space) wavelengths $\lambda$ or positive (angular) frequency $\omega$, where as usual $\omega = 2\pi c / \lambda$ for free-space phase velocity $c$.

We represent waves as complex functions in time, of form $\exp(i\omega t)$, knowing we can add the complex conjugate at the end for real waves. In this time window, an appropriate orthonormal Fourier basis set is

$$|\phi_{Ip}\rangle = \frac{1}{\sqrt{t_I}} \exp(i\omega_p t) \qquad (20)$$

where $\omega_p = 2\pi p / t_I$ with $p$ as a positive integer. The dimensionality of the input space here is then $M_{SI} \times M_I = M_I$.

For definiteness, we presume the number of output spatial modes $M_{SO}$ (e.g., different detectors or output fibers) is $M_{SO} = M_I$, as would be appropriate for a simple demultiplexer that put different input wavelengths in different output fibers or detectors. In the output space, we have to consider both wavelength and spatial aspects in the orthogonal basis functions, so the output space has

$$M_O = M_I M_{SO} = M_I^2 \qquad (21)$$

dimensions.

*Multicasting wavelength channels*

With no wavelength conversion, the most general relation, as in Eq. (1), between input and output functions, with the SVD as in Eq. (3), can be written in the form

$$
\begin{array}{cccc}
|\phi_O\rangle & V & D_{Diag} & U^\dagger & |\phi_I\rangle
\end{array}
$$

$$
\begin{bmatrix}
|\phi_{Ob1\lambda1}\rangle \\
|\phi_{Ob2\lambda1}\rangle \\
\vdots \\
|\phi_{Ob1\lambda2}\rangle \\
|\phi_{Ob2\lambda2}\rangle \\
\vdots \\
|\phi_{Ob1\lambda3}\rangle \\
|\phi_{Ob2\lambda3}\rangle \\
\vdots
\end{bmatrix}
=
\begin{bmatrix}
v_{11} & 0 & 0 & \cdots \\
v_{21} & 0 & 0 & \cdots \\
\vdots & \vdots & \vdots & \vdots \\
0 & v_{(M_I+1)2} & 0 & \cdots \\
0 & v_{(M_I+2)2} & 0 & \cdots \\
\vdots & \vdots & \vdots & \vdots \\
0 & 0 & v_{(2M_I+1)3} & \cdots \\
0 & 0 & v_{(2M_I+2)3} & \cdots \\
\vdots & \vdots & \vdots & \ddots
\end{bmatrix}
\begin{bmatrix}
s_{D1} & 0 & 0 & \cdots \\
0 & s_{D2} & 0 & \cdots \\
0 & 0 & s_{D3} & \cdots \\
\vdots & \vdots & \vdots & \ddots
\end{bmatrix}
\begin{bmatrix}
1 & 0 & 0 & \cdots \\
0 & 1 & 0 & \cdots \\
0 & 0 & 1 & \cdots \\
\vdots & \vdots & \vdots & \ddots
\end{bmatrix}
\begin{bmatrix}
|\phi_{I\lambda1}\rangle \\
|\phi_{I\lambda2}\rangle \\
|\phi_{I\lambda3}\rangle \\
\vdots
\end{bmatrix}
$$

(22)

Here we have written out vectors for the input function $|\phi_I\rangle$ in terms of its components $|\phi_{I\lambda_p}\rangle$ at wavelengths $\lambda_p$, and for the output function $|\phi_O\rangle$ in terms of its components $|\phi_{Obq\lambda_p}\rangle$ in output spatial mode (e.g., fiber or detector) $b_q$ at wavelength $\lambda_p$. In Eq. (22), note first that the matrix $U^\dagger$ is just an identity matrix. (We could permute the rows of $U^\dagger$ with corresponding permutations of the singular values and of elements of $V$, but this would be purely mathematical, leading to an entirely equivalent description of the same physical problem.) The reason for having no real choice of these rows is because of the "mask-based" nature of frequency filters without frequency conversion. In the matrix $V$, because there is no frequency conversion, a given input frequency only can be coupled to output modes of the same frequency, hence all the zeros in the columns of $V$.

In counting the number of numbers required to specify $V$ in the most general case here, to specify one column of $V$ we require $2M_I - 2$ real numbers ($2M_I$ to specify the $M_I$ generally non-zero complex elements, minus 1 because we arbitrarily fix the phase in each column, setting overall phases with the singular values, and minus 1 from normalization). These numbers essentially allow us to distribute arbitrary amounts of the input at a given wavelength among the various output spatial modes (e.g., fibers or detectors). So the total number of real numbers required to specify $\vee$ is $M_I(2M_I - 2)$. Adding the $2M_I$ real numbers to specify the singular values gives a total

$$N_D = 2M_I^2 \qquad (23)$$

If we do not care about the phases of any of the couplings from input to output, which would be the case if we were coupling into power detectors only, then we can drop the factor of 2 here (we have $M_I - 1$ phase factors in each of the $M_I$ columns of $\vee$ and $M_I$ phase factors in the singular values – a total of $M_I^2$ phase factors altogether, all of which we would neglect), leaving

$$N_D = M_I^2 \qquad (24)$$

If we only care about the signal power into each output spatial mode (and not the wavelengths), in communications a component with this functionality implements arbitrary multicasting (i.e., the ability to controllably route each input channel to any combination of output channels). Because there is no wavelength or frequency conversion, this device is not maximally functional, but it is maximally connected, mapping all $M_I$ input channels through the system. It may not be obvious how to make this device (at least without power splitting), but at least we have evaluated the minimum complexity required to make a device that can accomplish any such arbitrary division of input wavelengths among output spatial channels.

*Routing wavelength channels*

Suppose we simply want to route each incoming wavelength to a different output spatial mode, as in a wavelength router. For the $\vee$ matrix in Eq. (22), there will therefore only be one element (with amplitude 1) in each column of $\vee$ (and that element will be one of the non-zero elements of the columns in Eq. (22)).

To understand the number of real numbers we need to specify here, we need a slightly different approach from that used up till now, where the real numbers themselves have always been coefficients in the matrix elements. Now, the number we must specify is which row of a matrix column is non-zero. On the face of it, this row number is an integer, but an integer is a special case of a real number. As we will see below, specifying an integer here anyway means specifying a real number (e.g., the center wavelength of a resonator) in the physical device. To specify the matrix $\vee$, then, we specify an integer (the row number) for each column. Given that we are routing each wavelength to a unique output spatial mode, once we have specified this integer for each of $M_I - 1$ columns, we know what the integer must be for the final column (because it must correspond to the remaining unused output spatial mode). So, specifying $\vee$ for this router means specifying $M_I - 1$ real numbers.

Specifying this device in general therefore requires these $M_I - 1$ real numbers plus $2M_I$ real numbers to specify amplitude and phase of the singular values, giving

$$N_D = 3M_I - 1 \qquad (25)$$

If we do not care about the phase of the outputs, we can eliminate $M_I$ of these, leaving

$$N_D = 2M_I - 1 \qquad (26)$$

This still allows us to control output amplitudes. If we do not require separate control of amplitudes, then

$$N_C = M_I - 1 \qquad (27)$$

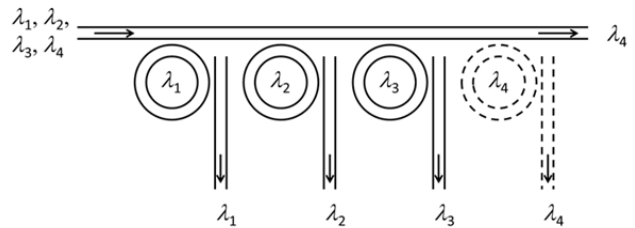

Fig. 4. Wavelength demultiplexing of four wavelengths $\lambda_1$, $\lambda_2$, $\lambda_3$, and $\lambda_4$ with ring resonator drop filters each of different radius to resonate at different ones of the input wavelengths. The fourth, dashed channel is optional in principle because only the fourth wavelength remains in the original guide with ideal devices.

Fig. 4 shows an example (and relatively standard) architecture for taking four wavelengths entering in one waveguide and separating them to different output waveguides. Ring resonators [47] operate as drop filters to separate the wavelengths (wavelengths $\lambda_1$, $\lambda_2$, $\lambda_3$, and $\lambda_4$ are not necessarily in any particular order here). The starting point for this category of devices is that we presume we have resonators to work with that have high enough finesse and large enough free spectral range that they can only pass one of the wavelengths of interest at a time, and with and equal input and output couplings (so they can be 100% transmitting on resonance). Then the only (real number) physical variable to be set is the ring radius (which sets the resonator center wavelength). To separate 4 wavelengths according to Eq. (27), we only need to set 3 real numbers, which is correct in this architecture. The first three rings (shown in solid lines in Fig. 4) separate off the first three wavelengths, leaving the fourth wavelength alone propagating in the original waveguide. As a practical matter, we might build a fourth drop filter (shown in dashed lines in Fig. 4) because we might not want to rely on the perfection of the first three in completely dropping their wavelengths, but the formula Eq. (27) correctly predicts the minimum complexity required here.

*General wavelength demultiplexing example*

A recent wavelength splitting device [42] designed for three different wavelengths used the lateral positions of 30 slits to design the device. This actual device was not designed to implement arbitrary multicasting – indeed it was only initially designed to implement a simple demultiplexer (see below) – but subsequent unpublished calculations [48] did verify that similar designs using this approach could achieve the arbitrary routing demultiplexer. Since it was only designed to split three wavelengths, the 30 slit positions should give enough variables in design to implement the multicasting router, which would only require 9 variables, according to Eq. (24).

*Simple demultiplexer*

A simple demultiplexer takes specific input frequencies and delivers them to specific progressive (and/or cyclic, as in a waveguide grating router) output positions. It does not put linear combinations of input frequencies at specific output positions, nor does it take one input frequency and distribute it to several outputs, and, other than possibly for the position of the first output, it does not allow selection of which other frequency goes to which other output spatial mode. In the matrix resulting $\vee$ as in Eq. (22), there is one "1" in each column, and these move progressively (and/or cyclically) through the available positions in each successive column. Other than possibly one variable to choose where the "1" is in the first column, we now have essentially no further choice. In such a simple demultiplexer, we may also not care about the phase of the individual outputs, and we may ask for no control over the amplitude (simply wanting it to be the largest the device can deliver). In that case, we have essentially only one real number variable to choose, which is the position of the "1" in the first column of $\vee$.

Of course, we know we can achieve this function with a simple grating. Possibly, we should regard ourselves as choosing where the first wavelength goes, which we could do by choosing one physical variable, the grating period, just as we have chosen the resonant frequency of each resonator in the wavelength router above. This one variable essentially corresponds to the position of the "1" in the first column of $\vee$.

If we take this simple demultiplexer approach when we actually want to make a router, as above, then we are externalizing the remaining routing functionality to the mechanical design of the output waveguide layout. With the "near" end of each of a set of optical fibers connected progressively to the outputs of the simple demultiplexer, we would need the equivalent of a "patch panel", choosing to which patch panel output port we connect the "far" end of each fiber. This would require $M_I - 1$ choices of integers to place the fiber output (the final fiber going into the only remaining output slot on the patch panel). In this case, it does not matter where the "first" output from the grating goes because the fiber attachments on the patch panel can handle any such choice. So, we are back to the same answer as Eq. (27) for our router, with all the designed functionality externalized to the mechanics.

## 5. Conclusions

In this paper, we have laid out a way of establishing the minimum complexity we need in the design of optical components. One preliminary conclusion we draw is that it is not possible to establish the complexity required to make or design a device for any one purpose. We cannot quantify the inherent complexity or functionality of a device; by "inherent" functionality, we mean what the device can do even if we do not design it to do that. But, we can establish the complexity required to design any one of a given set or "category" of devices from a particular starting point.

To quantify that complexity for linear optical devices, we have used an approach based on the SVD of the mathematical "device" operator that relates outputs to inputs. We have argued that, for any given problem, we can reduce the corresponding mathematical spaces to ones with finite dimensions; that reduction allows us to count complexity, establishing a "complexity number" – the number of real numbers we must specify to design any device within a given category. The core of the method is to use the SVD form to help count the number of independent numbers required to specify the device. In many cases, this number is much less than the total number of numbers in the matrix that represents the device operator.

We have defined concepts – "maximally functional", "maximally connected" and a "mode coupling number" – that help in categorizing results of this analysis. We have discussed several examples, including a single-mode to single-mode converter, more general mode-converters, and various wavelength filters, demultiplexers, and routers. We have examined a particularly broad and useful class that we call "mask-based" devices, which includes many common optical components such as lenses, mirrors, gratings, and fixed wavelength filters; these mask-based devices have, on the one hand, very useful inherent functionalities, thereby allowing lower design complexities than we might expect, and, on the other hand, substantial constraints on what functions they can ever be designed to perform no matter how complex we make them.

We have also examined examples of unconventional and novel devices that may be capable of any linear mapping between inputs and outputs within broad categories, and hence capable of functions that up till now have been difficult or impossible to design. An important conclusion of this work is that we can establish clear minimum bounds on the complexity required to allow the design of broad classes of optical devices, including such unconventional ones. We hope the results help especially in clarifying requirements and capabilities in optical devices, especially as we exploit emerging nanophotonic structures for new optical functionalities. We note, too, that the approach here is sufficiently general to apply to linear devices generally, including devices operating on a broad range of different kinds of waves and functions, and including spatial, temporal, and wavelength-dependent properties and functionalities.

# Appendix - Counting the number of input and output modes

If it is not immediately obvious what are the dimensionalities of the input and output spaces $H_I$ and $H_O$, we can establish those through the following procedure. This approach is based on establishing the so-called "communications modes" into and out of the device, in particular those with a coupling strength above some chosen threshold; because of the mathematics of this approach, for any finite threshold, the number of communications modes is also finite.

## A.1 Transmitting and receiving spaces

We consider two more mathematical spaces, the "transmitting" and "receiving" spaces, $H_T$ and $H_R$, respectively. $H_T$ is the space from which the input waves come. For example, it might be a scene of which we are taking a picture. $H_R$ is the space where we ultimately put the output of the device, such as the film or the detector array in a camera. The device itself might be the lens in the camera, with an input space $H_I$ that is describing the field on the front surface of the lens and an output space $H_O$ that is describing the field on the back surface of the lens (inside the camera).

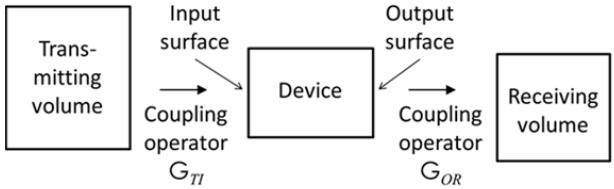

Fig. 5. Illustration of example transmitting, device, and receiving volumes. Sources or waves in the transmitting volume lead to waves on the device input surface through the coupling operator $G_{TI}$. Waves on the device output surface lead to waves in the receiving volume through the coupling operator $G_{OR}$.

Fig. 5 illustrates an example configuration. We have introduced two new operators, $G_{TI}$ and $G_{OR}$. $G_{TI}$ couples sources or waves $|\phi_T\rangle$ in the transmitting volume (or, more generally, in $H_T$) to functions $|\phi_I\rangle$ at the device input (or in $H_I$). Similarly, $G_{OR}$ couples functions $|\phi_O\rangle$ at the device output, (or in $H_O$) to functions $|\phi_R\rangle$ in the receiving volume (or in $H_R$). The operators $G_{TI}$ and $G_{OR}$ depend on the wave equations, the various spaces, and on how we set up the problem. Formally, for coupling source functions in one volume to resulting waves in the other, these operators are essentially the Green's functions of those wave equations. For coupling from waves on one surface to resulting waves on another, these operators are corresponding diffraction operators.

Whether we use volumes or surface in these various spaces depends on the problem; for example, in a camera looking at a three-dimensional scene, the transmission is from the scene volume, and if we are focusing the camera by moving the "film" plane, the receiving volume includes all possible positions of the film plane. Mathematically we could also use the entire device volume for both the input and output volumes of the device; if the device is some complicated volume scatterer, such an approach might be more complete mathematically than just considering fields at the "input" and "output" surfaces.

## A.2 Communications modes

The procedure for establishing the numbers of modes to use at the input ($M_I$) and output ($M_O$) is based on the SVD of the coupling operators $G_{TI}$ and $G_{OR}$. This decomposition establishes the so-called "communications modes" associated with these operators, and it is the counting of these that determines $M_I$ and $M_O$.

The idea of communications modes has been presented in Ref. [49] and this concept has seen various uses in optics and wireless communications [29, 50 - 56]. The SVD of the coupling operator $G_{TI}$ between the transmitting and device input spaces establishes sets of orthonormal functions $|\phi_{Tp}\rangle$ in the transmitting space $H_T$ and $|\phi_{Ip}\rangle$ in the device input space $H_I$ with associated coupling strengths (singular values) $s_{TIp}$. Formally, as in Eq. (2),

$$G_{TI} = \sum_p s_{TIp} |\phi_{Ip}\rangle\langle\phi_{Tp}| \qquad (28)$$

and, in the normal fashion for SVD,

$$G_{TI}^\dagger G_{TI} |\phi_{Tp}\rangle = |s_{TIp}|^2 |\phi_{Tp}\rangle \qquad (29)$$

$$G_{TI} G_{TI}^\dagger |\phi_{Ip}\rangle = |s_{TIp}|^2 |\phi_{Ip}\rangle \qquad (30)$$

We can construct a similar set of equations for the SVD

$$G_{OR} = \sum_q s_{ORq} |\phi_{Rq}\rangle\langle\phi_{Oq}| \qquad (31)$$

of $G_{OR}$, leading to orthonormal functions $|\phi_{Oq}\rangle$ in the device output space $H_O$ and $|\phi_{Rq}\rangle$ in the device input space $H_R$ with associated coupling strengths (singular values) $s_{ORq}$.

Just as in the SVD of the device operator $D$, these SVDs can essentially always be performed; the key requirement is that the operators $G_{TI}$ and $G_{OR}$ are mathematically "compact", as are typical operators in wave problems (see [32] for a discussion of compactness). The resulting sets of functions corresponding to non-zero singular values are complete in the same sense we are using for the "mode converter" basis sets $|\phi_{DOm}\rangle$ and $|\phi_{DIm}\rangle$ above.

## A.3 Counting communications modes

In wave problems involving free space propagation, the various sets here – $|\phi_{Tp}\rangle$ and $|\phi_{Ip}\rangle$ for $G_{TI}$ and $|\phi_{Oq}\rangle$ and $|\phi_{Rq}\rangle$ for $G_{OR}$ – are generally infinite, so we have not yet generally established finite numbers for $M_I$ and $M_O$. However, the singular values obey a sum rule [49], which can resolve this problem.

Quite generally, we can define sums $S_{TI}$ and $S_{OR}$ obeying the rules

$$\sum_p |s_{TIp}|^2 = Tr(G_{TI}^\dagger G_{TI}) = S_{TI} \quad (32)$$

and similarly

$$\sum_q |s_{ORq}|^2 = Tr(G_{OR}^\dagger G_{OR}) = S_{OR} \quad (33)$$

The proof of these sum rules follows simply if we note that (i) the trace $Tr$ (i.e., sum of the diagonal elements) of an operator is independent of the (complete) basis used to represent it and (ii) one such basis is the eigenbasis ($|\phi_{Tp}\rangle$ for $G_{TI}^\dagger G$ or $|\phi_{Oq}\rangle$ for $G_{OR}^\dagger G_{OR}$), for which the matrix is diagonal with diagonal elements $|s_{TIp}|^2$ for $G_{TI}^\dagger G$ or $|s_{ORq}|^2$ for $G_{OR}^\dagger G_{OR}$.

Because the trace is independent of the basis, we can typically evaluate $S_{TI}$ and $S_{OR}$ using continuous (i.e., delta-function) basis sets [49], which means performing some volume and/or surface integrals. For example, for a simple scalar wave equation with Green's function $G_{TI}(\mathbf{r}_T, \mathbf{r}_I)$, $G_{TI} \equiv G_{TI}(\mathbf{r}_T, \mathbf{r}_I)$ where $\mathbf{r}_T$ and $\mathbf{r}_I$ are position vectors in the transmitting and device input volumes or surfaces, $V_T$ and $V_I$ respectively. Then

$$S_{TI} = Tr(G_{TI}^\dagger G_{TI}) = \int_{V_T} \int_{V_I} |G_{TI}(\mathbf{r}_T, \mathbf{r}_I)|^2 dV_I dV_T \quad (34)$$

and similarly for $S_{OR} = Tr(G_{OR}^\dagger G_{OR})$.

We can reasonably decide that there are some minimum connection strengths $|s_{TImin}|^2$ and $|s_{ORmin}|^2$ of interest; connections below these strengths we can consider to be so low that the coupling is negligible or insufficiently useful to us in the device. We then know immediately that we cannot usefully have more than $S_{TI}/|s_{TImin}|^2$ and $S_{OR}/|s_{ORmin}|^2$ communications modes at the device input and output respectively. More stringent limits can be obtained if we solve the SVDs of Eqs. (28) and (31). Then, by progressively adding up an ordered list of the $|s_{TIp}|^2$ and $|s_{ORq}|^2$, for example, from largest on downwards, we can decide when there is insufficient capacity left in the sum rules for any further couplings that are strong enough to be worth considering (e.g., when the ordered sum is within $|s_{TImin}|^2$ or $|s_{ORmin}|^2$ of the totals $S_{TI}$ and $S_{OR}$, respectively); at that point in each case we can truncate the basis sets to obtain practical numbers of basis functions $M_I$ (from the $S_{TI}$ comparison) and $M_O$ (from the $S_{OR}$ comparison), respectively. (See Ref. [49] for examples of such convergence.)

In a broad range of optical situations, such as paraxial optics between plane-parallel surfaces or plane-parallel volumes of uniform thickness and of transverse dimensions that are many wavelengths in size, the behavior of the $|s_{TIp}|^2$ and $|s_{ORq}|^2$ can become particularly simple: Up to a specific number in each case (essentially, the $M_I$ and $M_O$ we will want to choose in each case), the coupling strengths $|s_{TIp}|^2$ and $|s_{ORq}|^2$ are each approximately constant independent of the index $p$ or $q$ respectively, and then they drop off dramatically once we try to pass the normal diffraction limit. This case is analyzed in detail in Ref. [49].

Note, incidentally, that this approach is not restricted to purely spatial problems; it can be used in the time or frequency domain as well or in combinations of spatial and temporal (or frequency) domains or with other degrees of freedom (such as polarization or spin). It works provided only that the coupling operators are compact. (In time-dependent problems, it may be necessary to impose finite frequency bandwidths; otherwise time derivative operators are generally not bounded and therefore not compact.)

## Acknowledgements

This project was supported by funds from Duke University under an award from DARPA InPho program, and by the AFOSR Robust and Complex On-Chip Nanophotonics MURI.